\documentclass[prb,aps,twocolumn,floatfix,showpacs,amsmath,amssymb]{revtex4}
\usepackage{graphics}
\usepackage{graphicx}
\usepackage[dvips]{color}
\begin{document}
%\title{Environmentally-Induced Rabi Oscillations \\ and Decoherence in Superconducting Qubits}
\title{Enhancing decoherence times in superconducting qubits via
circuit design}
\author{Kaushik Mitra$^1$ and C. A. R. S{\'a} de Melo$^2$}
\affiliation{1. Joint Quantum Institute and Department of Physics \\ 
University of Maryland College Park MD 20742}
\affiliation{2. School of Physics, 
Georgia Institute of Technology, Atlanta, GA 30332}

\date{\today}

\begin{abstract}
We study decoherence effects in qubits coupled to environments that exhibit 
resonant frequencies in their spectral function. We model the coupling of the
qubit to its environment via the Caldeira-Leggett formulation of 
quantum dissipation/coherence, and study the simplest example of 
decoherence effects in circuits with resonances such as a dc SQUID phase qubit 
in the presence of an isolation circuit, which is designed to enhance 
the coherence time. We emphasize that the spectral density of the environment
is strongly dependent on the circuit design, and can be engineered 
to produce longer decoherence times.
We begin with a general discussion of superconducting
qubits such as the flux qubit, the Cooper pair box and the phase qubit 
and show that in these kinds of systems appropriate circuit design can greatly 
modify the spectral density of the environment and lead to enhancement
of decoherence times. 
In the particular case of the phase qubit, for instance, 
we show that when the frequency of the qubit is at least
two times larger than the resonance frequency of the environmental 
spectral density, the decoherence time of the qubit is a few orders 
of magnitude larger than that of the
typical ohmic regime, where the frequency of the qubit is much smaller 
than the resonance frequency of the spectral density. 
In addition, we demonstrate that the environment does not only affect the decoherence
time, but also the frequency of the transition itself, which is shifted from 
its environment-free value. 
Second, we show that when the qubit frequency is nearly the same as the
resonant frequency of the environmental spectral density, 
an oscillatory non-Markovian decay emerges, as the qubit and its 
environment self-generate Rabi 
oscillations of characteristic time scales shorter than the decoherence time. 
\end{abstract}
\maketitle

\section{Introduction}
\label{sec:introduction}

The possibility of using quantum mechanics to manipulate information 
efficiently has lead, through advances in technology, to the plausibility of building a quantum 
computer using
two-level systems, also called quantum bits or qubits. 
Several schemes have been proposed as attempts to manipulate qubits in atomic, molecular and 
optical physics (AMO),
and condensed matter physics (CMP), however it is still very difficult to implement 
a scheme that gives both long decoherence times and is scalable. 

In AMO, the most promising schemes are 
trapped ion systems~\cite{monroe-95}, and ultracold atoms in
optical lattices~\cite{brennen}.
On the CMP side, the pursuit of solid state qubits has been 
quite promising in spin systems~\cite{hanson, hayashi} and 
superconducting devices~\cite{devoret-02,lobb-01,shnirman-97}. While the manipulation
of qubits in AMO has relied on the existence of qubits in a lattice of ions or 
ultra-cold atoms and the use
of lasers, the manipulation of qubits in CMP has relied on the 
NMR techniques (spin qubits) and the Josephson
effect (superconducting qubits). Integrating qubits into a full quantum
computer requires a deeper understanding of decoherence effects in a single qubit
and how different qubits couple. 

In AMO systems Rabi oscillations in single qubits have been observed over time scales 
of miliseconds since each qubit can be made quite isolated 
from its environment~\cite{monroe-95}, however it has been 
very difficult to implement 
multi-qubit states as the coupling between different qubits is not yet fully controllable. 
On the other hand, in superconducting qubits Rabi oscillations
have been observed~\cite{devoret-02, martinis-05} 
over shorter time scales (500ns), since these qubits are coupled to many environmental 
degreees of freedom, and require very careful circuit design. 
For superconducting qubits, it has been shown experimentally~\cite{martinis-05} 
that sources of decoherence from two-level states within the insulating barrier of 
a Josephson junction can be significantly reduced by using 
better dielectrics and fabricating junctions of small area $(\lesssim 10 {~\mu \rm m^2})$.

In this manuscript, we use the Caldeira-Leggett formulation of quantum dissipation
to analyze decoherence effects of generic superconducting qubits coupled to environments 
that exhibit a resonance in their spectral density. This is an extension of our previous work~\cite{mitra-09},
where the general idea presented in this paper 
in the context of dc-SQUID phase qubits.
Here, we show first how the characteristic 
spontaneous emission (relaxation) lifetimes $T_1$ for
flux~\cite{mooij-99, van-der-wal-03}, phase~\cite{martinis-02} and charge~\cite{koch-07} qubits 
can be substantially enhanced, 
when each of them is coupled to an environment with a resonance,
provided that the frequency of operation of the qubit is about twice as large
as the frequency of the environmental resonance.
Second, we show that the coupling to the environment does not only cause decoherence, 
but also changes the qubit frequency.
Lastly, we show that when the qubit frequency is nearly the same as the
resonant frequency of the environmental spectral density, 
an oscillatory non-Markovian decay emerges, as the qubit and its 
environment self-generate Rabi oscillations of characteristic time 
scales shorter than the decoherence time.

The paper is organized as follows. In Sec.~\ref{sec:spectral-densities}, 
we derive the environmental spectral density
for flux, phase, and charge qubits when each
of them is coupled to an environment with a resonance. 
In Sec.~\ref{sec:bloch-redfield-equations}, we use the Caldeira-Leggett formulation 
of quantum dissipation to derive the Bloch-Redfield equations 
and to calculate the decoherence and relaxation times of these qubits.
There, for the particular case of the phase qubit, we show that when 
the frequency of the qubit is at least two times larger than the resonance 
frequency of the environmental spectral density, the decoherence time 
of the qubit is a few orders of magnitude larger than that of the
typical ohmic regime, where the frequency of the qubit is much smaller 
than the resonance frequency of the spectral density. 
In Sec.~{\ref{sec:frequency-normalization}, we calculate the renormalization 
of the qubit frequency due to dressing of the two-level
system with environmental degrees of freedom. 
In Sec.~{\ref{sec:non-markovian}, we derive an essentially exact non-perturbative master equation, 
when these systems are near resonance and where the rotating wave approximation can be applied. 
Our results indicate that when the qubit frequency is nearly the same as the
resonant frequency of the environmental spectral density, 
an oscillatory non-Markovian decay emerges, as the qubit and its 
environment self-generate Rabi oscillations of characteristic time scales 
shorter than the decoherence time. Finally, we present our conclusions 
in Sec.~\ref{sec:conclusions}.

\section{Spectral densities of superconducting qubits with environmental resonances}
\label{sec:spectral-densities}

In this section, we discuss three examples of superconducting
qubits coupled to environments (circuits) which have resonances in their 
spectral functions. We begin our discussion with a simple 
flux qubit, then move to a charge qubit and finally to a phase qubit.
\begin{figure}[htb]
\centerline{
\scalebox{0.40}{
\includegraphics{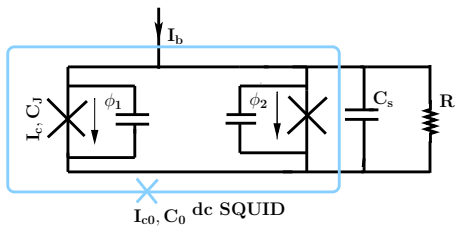}
}
}
\caption{
(Color-online) Flux qubit measured by a dc-SQUID gray (blue) line. 
The qubit corresponds to the inner 
SQUID loop with critical current $I_c$ and capacitance $C_J$ 
for both Josephson junctions denoted by the large $\times$ symbol. 
The inner SQUID is shunted by a capacitance $C_s$, and environmental resistance $R$ 
and is biased by a ramping current $I_b$. The dc-SQUID loop has 
junction capatitance $C_0$ and critical current $I_{c0}$.}
\label{fig:one}
\end{figure}

In Fig.~\ref{fig:one}, we show a flux qubit (inner-loop), which is measured by a 
dc-SQUID (outer loop). To study the decoherence and relaxation time scales in such
a system it is necessary to understand how noise is transferred from from the dc-SQUID
to the qubit. 

For the circuit displayed in Fig.~\ref{fig:one},
the classical equation of motion for the dc-SQUID (outer-loop) is
\begin{equation}
\label{eqn:ceq1}
C_0 \ddot \phi   + \frac{2\pi}{\Phi_0}I_{c0}\sin\phi -\frac{2\pi}{\Phi_0}I + 
\int_0^t dt'  Y(t - t') \dot\phi (t') = 0
\end{equation}
where $\phi$ is the gauge invariant phase across the Josephson junction of the dc-SQUID, 
$I_{c0}$ is the critical current of its junction, $\Phi_0 = h/2e$ is the flux quantum. 
The last term in Eq.~(\ref{eqn:ceq1}) is the dissipation term due to effective admittance
$Y(\omega)$ felt by the outer dc-SQUID. 
In this case the total induced current in the dc-SQUID (outer loop) is~\cite{tian-02} 
\begin{equation}
\label{eqn:I}
I = \left(\frac{\Phi_0}{2\pi}\right)\frac{4}{L_{dc}} \langle\delta \phi_0 \sigma_z \rangle + \left(\frac{\Phi_0}{2\pi}\right)\frac{4}{L_{dc}}\langle\phi_m\rangle +
\left(\frac{2\pi}{\Phi_0}\right) J_1 \langle\phi_p\rangle.
\end{equation}
Here $\phi_p = (\phi_1 + \phi_2)$ and $\phi_m = (\phi_2 - \phi_1)$ 
are the sum and difference of the
gauge invariant phases $\phi_1$ and $\phi_2$ across the junctions of 
the inner SQUID, $L_{dc}$ is the self-inductance of the inner SQUID,
and $J_1$ is the bilinear coupling between $\phi_m$ and $\phi_p$ at the potential energy minimum.
The term $\delta\phi_0 = \pi M_q I_{cir} / \Phi_0$, where $I_{cir}$ is the circulating current 
of the localized states of the qubit (described in terms of Pauli matrix $\sigma_z$), 
and $M_q$ is the mutual inductance between the qubit and the outer dc-SQUID.

For the charge qubit coupled to a transmission line
resonator~\cite{blais-04,koch-07} shown in Fig.~\ref{fig:two}, the classical equation of motion 
for the charge Q is 
\begin{eqnarray}
\label{eqn:ceq2}
V_g(\omega) =\left(-\frac{\omega^2 L_J(\omega)}{1-\omega^2 L_J(\omega) C_J} + \frac{1}{C_g} 
+ i\omega Z(\omega)\right)Q(\omega)
\end{eqnarray}
Here,  $V_g$ is the gate voltage, $C_g$ is the gate capacitance, $L_J$ and $C_J$ 
are the Josephson inductance and capacitance respectively, $Z(\omega)$ is the effective impedance 
seen by the charge qubit due to a transmission line resonator (cavity), and $Q$ is the charge across $C_g$.  
\begin{figure}[htb]
\centerline{
\scalebox{0.30}{
\includegraphics{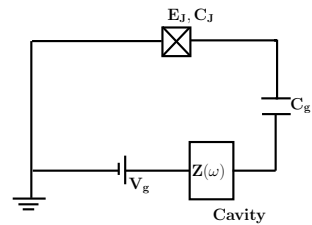}
}
}
\caption{Circuit diagram of the Cooper-pair box. The superconducting island 
(large $\times$)
is connected to a large reservoir through a 
Josephson junction with Josephson energy $E_J$ and capacitance $C_J$. 
The voltage bias $V_g$ is provided through a resonator (cavity) having
environmental impedance $Z(\omega)$, which is connected 
to the gate capacitance $C_g$ as shown.
}
\label{fig:two}
\end{figure}

In Fig.~\ref{fig:three}, we show the circuit for a phase qubit corresponding to an
asymmetric dc-SQUID~\cite{martinis-02, martinis-03}.
The circuit elements inside the dashed box form an isolation network which serves two purposes:
a) it prevents current noise from reaching the qubit junction; 
b) it is used as a measurement tool.

The classical equation of motion for the phase qubit 
shown in Fig.~\ref{fig:three} is 
\begin{equation}
\label{eqn:ceq3}
C_0 \ddot \gamma   + \frac{2\pi}{\Phi_0}I_{c0}\sin\gamma -\frac{2\pi}{\Phi_0}I + 
\int_0^t dt'  Y(t - t') \dot\gamma (t') = 0,
\end{equation}
where $C_0$ is the capacitance, $I_{c0}$ is the critical current, and $\gamma$ is the phase difference across 
the Josephson junction $J$ (large $\times$ in Fig.~\ref{fig:three}), 
while $I$ is the bias current, and $\Phi_0 = h/2e$ is the flux quantum. 
The last term of 
Eq.~(\ref{eqn:ceq3}) can be written as
$i\omega Y(\omega) \gamma (\omega)$ in Fourier space.
The admittance function  $Y(\omega)$ can be modeled as two additive
terms $Y(\omega) = Y_{iso} (\omega) + Y_{int} (\omega)$. 
The first contribution $Y_{iso} (\omega)$ is the admittance that results when a transmission
line of characteristic impedance $R$ is attached to the isolation junction
(here represented by a capacitance $C$ and a Josephson inductance $L$)
and an isolation inductance $L_1$. Thus,
$
Y_{iso} (\omega) =  Z_{iso}^{-1} (\omega) 
$
where $Z_{iso} (\omega) = (i\omega L_1) + \left[  R^{-1} + i\omega C + (i\omega  L)^{-1} \right]^{-1}$
is the impedance of the isolation network shown in Fig.~\ref{fig:three}. 
The replacement of the isolation junction by an LC circuit is justified because under
standard operating conditions the external flux $\Phi_{\rm a}$ varies to cancel the current flowing through 
the isolation junction making it zero biased~\cite{martinis-02}. 
Thus, the isolation junction behaves as a harmonic oscillator with
inductance $L$ which is chosen to be much smaller than $L_1$. 
The second contribution $Y_{int} (\omega)$ is an internal admittance 
representing the local environment of the qubit junction, such as defects in the oxide 
barrier, quasiparticle tunneling, or the substrate, and can be modeled by 
$Y_{int} (\omega) =  (R_0 + i\omega L_0)^{-1}$, where $R_0$ is the resistance
and $L_0$ is the inductance of the qubit as shown in Fig.~\ref{fig:three}.
\begin{figure}[htb]
\centerline{
\scalebox{0.42}{
\includegraphics{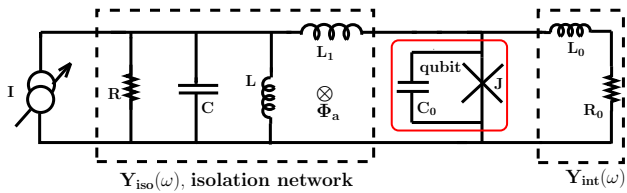}
}
}
\caption{(Color-online) 
Schematic drawing of a phase qubit with an RLC isolation circuit.
The phase qubit is shown inside the solid (red) box, the RLC isolation circuit
is shown inside the dashed box to the left, and the internal admittance circuit
is shown inside the dashed box to the right.
}
\label{fig:three}
\end{figure}

The equations of motion described in Eqs.~(\ref{eqn:ceq1}), (\ref{eqn:ceq2}), and (\ref{eqn:ceq3}),  
can be all approximatelly described by the effective spin-boson Hamiltonian 
\begin{equation}
\label{eqn:hamiltonian}
\widetilde{H} = \frac{\hbar \omega_{01}}{2}\sigma_z
+ \sum_k \hbar \omega_k b_k^\dagger b _k
+ H_{SB},
\end{equation}
written in terms of Pauli matrices $\sigma_i$ (with $i = x, y, z$) and boson 
operators $b_{k}$ and $b_{k}^\dagger$.
The first term in Eq.~(\ref{eqn:hamiltonian}) represents a two-level approximation
for the qubit (system) described by states $\vert 0 \rangle$ and $\vert 1 \rangle$ 
with energy difference $\hbar \omega_{01}$.
The second term corresponds to the isolation
network (bath) represented by a bath of bosons, where 
$b_{k}$ and $b_{k}^\dagger$ are the annihilation and creation operator of 
the $k$-th bath mode with frequency $\omega_{k}$. 
The third term is the system-bath (SB) Hamiltonian which corresponds to the coupling 
between the environment and the qubit. 

At the charge degeneracy point for the charge qubit (gate charge $N_g=0$),
at the flux degeneracy point (external flux $\Phi_{\rm ext}=\pi\Phi_0$) 
for the flux qubit, and at
the suitable flux bias condition for the phase qubit 
(external flux $\Phi_{\rm a}= L_1\phi_0$)
$H_{SB}$ reduces to
\begin{equation}
\label{eqn:system-bath-hamiltonian}
H_{SB} = 
\frac{1}{2}\sigma_x\hbar\langle 1\vert v\vert 0\rangle 
\sum_k \lambda_{k1} \left( b_{k}^\dagger +  b_{k} \right)
\end{equation}
where $v = \phi$ for the flux qubit, 
$v = Q$ for the charge qubit, and $ v = \gamma$ for 
the phase qubit. 
The spectral density of the bath modes 
\begin{equation}
J(\omega) = 
\hbar \sum_k \lambda_k^2  \delta \left( \omega- \omega_{k} \right)
\end{equation}
has dimensions of energy and can be written as 
\begin{equation}
\label{eqn:spectral-density-flux-phase}
J(\omega) = \omega {\rm Re} Y (\omega) 
\left(
\frac{\Phi_0}{2\pi}
\right)^2
\end{equation}
for flux and phase qubits or as
\begin{equation}
\label{eqn:spectral-density-charge}
J(\omega) 
= 
2 \hbar\omega {\rm Re} Z (\omega) 
\frac{e^2}{\hbar}
\end{equation}
for charge qubits.

For the flux qubit circuit shown in Fig.~\ref{fig:one}, 
the shunt capacitance $C_s$ is used to control the environment, 
while the Ohmic resistance of the circuit is modelled by $R$. 
In this case, the environmental spectral density is~\cite{tian-02}
\begin{equation}
\label{eqn:spectral-density}
J_{1}\left(\omega\right) =
\frac{\alpha_1\omega}{\left(1-\omega^2/\Omega_1^2\right)^2+4\omega^2\Gamma_1^2/\Omega_1^4}.
\end{equation}
when $\omega_m \gg {\rm max}(\Omega_1, \omega_{01})$, and 
%
%% From Carlos: define \omega_p and \omega_0 are.
%
%
%%
%%
%
when the dc-SQUID is far away from the switching point 
to be modelled by an ideal inductance $L_J$. The inner
oscillator frequency $\omega_m = \sqrt{2/L_{dc}C_J}$
and the qubit frequency is $\omega_{01}$.
}
The external 
oscillator frequency 
\begin{equation}
\Omega_1
= 
\sqrt{2\pi  I_c^{\rm eff}/C_s \Phi_0}[1-(I_b/I_c^{\rm eff})^2]^{1/4},
\end{equation} 
and plays the role of
the resonant frequency, 
where $I_{c}$ is the critical current for each of two Josephson junctions. 
Also, $\Gamma_1=  {1}/(C_s R)$ corresponds to the resonance width, and
\begin{equation}
\alpha_1  
=  
\frac{2 (e I_{cir} I_b M_q)^2}{C_s^2 \hbar^2 R \Omega_1^4}
\end{equation}
reflects the low frequency behavior. 
The coupling between the flux qubit and the outer dc-SQUID emerges from
the interaction of the persistent current $I_{cir}$ of the qubit 
and the bias current $I_b$ of the dc-SQUID via their mutual inductance $M_q$.

The spectral density for the charge qubit shown in Fig.~\ref{fig:two}
is obtained 
via Eq.~(\ref{eqn:spectral-density-charge}) by determining
the real part of the impedance $Z(\omega)$. In this case, we 
need to solve for the normal modes of the resonator and transmission lines, 
including an input impedance
$R$ at each end of the resonator. This procedure results in the  
spectral density~\cite{blais-04}
\begin{equation}
\label{eqn:spectral-density1}
J_{2}\left(\omega\right) 
=
\frac{e^2\Omega_2}{\ell c}
\frac{\Gamma_2}{(\omega-\Omega_2)^2+(\Gamma_2/2)^2},
\end{equation}
where $\Omega_2$ is the resonator frequency, $\ell$ is resonator length, 
$c$ is the capacitance 
per unit length of the transmission line. 
The quantity $\Gamma_2 = \Omega_2/Q$ where $Q$ is the quality 
factor of the cavity. 

For the phase qubit shown in Fig.~\ref{fig:three}, the spectral density
$J(\omega)$ is given by Eq.~(\ref{eqn:spectral-density-flux-phase}),
and can be written in the compact form
$J(\omega) = J_{iso} (\omega) + J_{int} (\omega)$. The spectral
density of the isolation network is 
\begin{equation}
\label{eqn:spectral-density-isolation}
J_{iso}\left(\omega\right) = \left(\frac{\Phi_0}{2\pi} \right)^2
\frac{\alpha\omega}{\left(1-\omega^2/\Omega^2\right)^2+4\omega^2\Gamma^2/\Omega^4},
\end{equation}
where $\alpha=L^2/\left[ (L+L_1)^2 R \right] \approx (L/L_1)^2/R$ is the leading order term in the 
low frequency ohmic regime, 
\begin{equation}
\Omega
=
\sqrt{
\frac {(L+L_1)}{LL_1C}
} 
\approx 
\frac{1}{\sqrt{LC}}
\end{equation}
is essentially the resonance frequency, and 
$\Gamma=1/(2CR)$ plays the role of resonance width. 
Here, we used $L_1 \gg L$ corresponding to the relevant experimental regime.
%
%%
%\textcolor{red}{*** Check for the dimensions of $J_{iso}$ and $J_{int}$ ***}
%%
%
Notice that $J_{iso} (\omega)$ has Ohmic behavior at low frequencies
since
\begin{equation}
\lim_{\omega\rightarrow 0}
\frac{J_{iso} (\omega)}{\omega} 
= 
\left(
\frac{\Phi_0}{2\pi}
\right)^2 
\left(
\frac{L}{L_1}
\right)^2
\frac{1}{R},
\end{equation}
but has a peak at frequency $\Omega$ with broadening controlled by $\Gamma$.
In addition, notice that the parameter
\begin{equation}
\frac{\Gamma}{\Omega^2}
 = 
\frac{LL_1}{\left(2R\left(L_1+L\right)\right)} 
\approx
\frac{L}{R}
\end{equation}
is independent of $C$. Therefore, when there is no capacitor ($C \to 0$), 
the resonance disappears and 
\begin{equation}
\label{eqn:sd-drude}
J_{iso} ( \omega ) =   \left(\frac{\Phi_0}{2\pi} \right)^2 \frac{\alpha\omega}{1+4\omega^2\Gamma^2/\Omega^4},
\end{equation}
reduces to a Drude term with characteristic frequency 
$\Omega^2/2\Gamma \approx  R/L$. 
The internal spectral density of the qubit 
\begin{equation}
\label{eqn:spectral-density-internal}
J_{int}\left(\omega\right) = \left(\frac{\Phi_0}{2\pi} \right)^2
\frac {(\omega/R_0)} {1 + \omega^2 L_0^2/R_0^2}
\end{equation}
is a Drude term with characteristic frequency $R_0/L_0$.
Notice that $J_{int} (\omega)$ also has Ohmic behavior at low frequencies
since
\begin{equation}
\lim_{\omega\rightarrow 0} 
\frac{J_{int} (\omega)}{\omega}
= 
\left(
\frac{\Phi_0}{2\pi}
\right)^2
\frac{1}{R_0}.
\end{equation}

Having discussed the spectral functions of the environment of three different
types of superconducting qubits and their corresponding circuits, we move next
to the discussion of the decoherence properties within the Bloch-Redfield 
description of the time evolution of the density matrix.

\section{Bloch-Redfield equations: decoherence properties}
\label{sec:bloch-redfield-equations}

In this section, we investigate the 
relaxation $T_1$ and decoherence $T_2$ times,  
for the three types of superconducting qubits (flux, charge and phase) 
described in section~\ref{sec:spectral-densities}.
In order to obtain $T_1$ and $T_2$, we write the 
Bloch-Redfield equations~\cite{weiss-99}
\begin{equation}
\label{eqn:d-matrix}
\dot{\rho}_{nm} = -i\omega_{nm}\rho_{nm} + \sum_{kl} R_{nmkl}\rho_{kl}
\end{equation}
for the density matrix $\rho_{nm}$ of the spin-boson Hamiltonian in 
Eq.~(\ref{eqn:hamiltonian}) and~(\ref{eqn:system-bath-hamiltonian})
derived in the Born-Markov limit.
Here all indices take the values $0$ and $1$ corresponding to the
ground and excited states of the qubit, respectively, while $\omega_{nm}=(E_n-E_m)/\hbar$
is the frequency difference between states $n$ and $m$.
The Redfield rate tensor is
\begin{equation}
R_{nmkl} = -\Gamma_{lmnk}^{(1)}-\Gamma_{lmnk}^{(2)} + \delta_{nk} \Gamma_{lrrm}^{(1)}  
 + \delta_{lm} \Gamma_{nrrk}^{(2)},  
\label{eq:redfield-tensor}
\end{equation}
where repeated indices indicate summation, and
\begin{eqnarray}
\Gamma_{lmnk}^{(1)} = 
\hbar^{-2}  \int_0^\infty dt   
e^{-i\omega_{nk} t} \langle H_{SB,lm}(t) H_{SB,nk}(0)\rangle, \\
\Gamma_{lmnk}^{(2)} = 
\hbar^{-2}  \int_0^\infty dt   
e^{-i\omega_{lm} t} \langle H_{SB,lm}(0) H_{SB,nk}(t)\rangle. 
\end{eqnarray}
Under these conditions, the relaxation rate
becomes
\begin{equation}
\label{eqn:t1}
\frac{1}{T_1} =  -\sum_n\limits R_{nnnn}
= \frac{J(\omega_{01})}{M\omega_{01}}
\coth \left( \frac{\hbar\omega_{01}}{k_BT} \right), 
\end{equation}
where we used the transition probability 
$|\langle 0|\nu|1\rangle|^2 = \left[ M\omega_{01} \right]^{-1}$
to define the variable $M$, which has dimensions of mass $\times$ 
area (or energy $\times$ squared-time) 
and is refered to as the {\it mass} of the qubit. 
For the flux and phase qubits, the \it mass} is 
$M\equiv \left(\Phi_0/2\pi\right)^2C_0$,
where $C_0$ is the capacitance of the qubit. 
For the charge qubit, the {\it mass} is $M \equiv \hbar e/I_0$, 
where $I_0$ is the critical current. 
The frequency $\omega_{01}$ is the qubit frequency. 

The physical interpretation of $T_1^{-1}$ is as follows. For the system 
to make a transition it needs to exchange 
energy $E = \hbar \omega_{01}$ with the environment using a single boson. 
The factor $ \coth (\hbar \omega_{01}/k_B T) =   n(\omega_{01})+ 1 + n(\omega_{01})$ 
captures the sum of the rates for emission 
(proportional to $n(\omega_{01})+ 1$) and 
absorption (proportional to $n(\omega_{01})$ of 
a boson), where 
$n(\omega_{01}) = \left[ \exp (\hbar \omega_{01}/k_B T) - 1 \right]^{-1}$ 
is the Bose function.

\begin{figure}
\centerline{ \scalebox{0.7} {\includegraphics{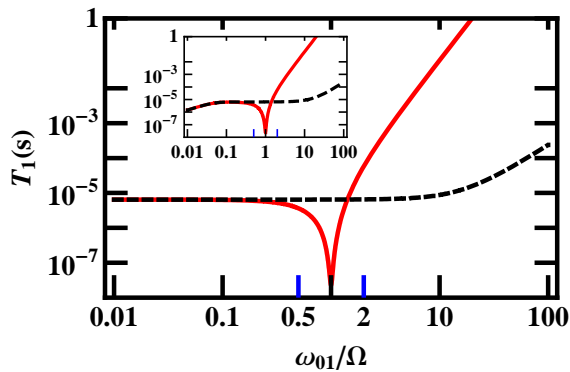}} } 
\caption{\label{fig:four}
(Color-online) $T_1$ (in seconds) as a function of qubit frequency 
$\omega_{01}$. The solid (red) curves describe the phase qubit 
with RLC isolation network (Fig.~\ref{fig:three}) with
parameters $R = 50$~ohms, $L_1=3.9$nH, $L=2.25$pH, $C=2.22$pF, and 
qubit parameters $C_0=4.44$pF,  $R_0 = \infty$ and $L_0 = 0$. 
The dashed curves correspond to an RL isolation network with 
the same parameters, except that $C = 0$. Main figure ($T =0$), 
inset ($T = 50$mK) with $\Omega = 141 \times 10^9$ rad/sec.
}
\end{figure}

Similarly, the decoherence rate given by the {\it off-diagonal} elements of the reduced 
density matrix $\rho$ is
\begin{equation}
\label{eq:ttwo}
\frac{1}{T_2} = {\rm Re} (R_{nmnm}) 
=\frac{1}{2T_1}+\frac{1}{T_\phi},
\end{equation}
where the dephasing rate 
\begin{eqnarray}
\frac{1}{T_\phi}
 & = & 
|\langle 0|\nu|0\rangle-\langle 1|\nu|1\rangle|^2 
\mathop{\lim }\limits_{\omega \to 0}
\frac{J(\omega)}{\omega}2k_B T
\end{eqnarray} 
This contribution originates from 
dephasing processes which randomize the phases
while keeping energy constant, i.e.,  transitions from a state into
itself. Hence, they exchange zero energy with the environment and $J(0)$ enters. The prefactor measures which fraction
of the total environmental noise leads to fluctuations 
of the energy splitting, such that only the components of noise which
are {\em diagonal} in the basis of energy eigenstates leads to pure dephasing~\cite{wilhelm-06}.
This effect, is thus complementary to that of the {\it off-diagonal} 
transition matrix element $\langle 0 \vert \nu \vert 1 \rangle$ which appears
in $T_1$. 
The zero frequency argument is a consequence of the Markov approximation. 
%More physically,it can be understood as a limiting procedure involving the duration of the experiment, 
%which converges 
%to $S(0)$ under the motional narrowing condition. 
%
%
Typically, $T_\phi \gg T_1$. This can be seen for example, in the Hamiltonian of the phase qubit, where  
$\langle0|\gamma|0\rangle\sim\langle1|\gamma|1\rangle \sim \partial^3 U/\partial \gamma^3$ is the cubic correction to the potential of the phase qubit.

In Fig.~\ref{fig:four}, $T_1$ is plotted for the phase qubit
as a function of the qubit frequency $\omega_{01}$ in the
case of spectral densities describing an RLC 
[Eq.~(\ref{eqn:spectral-density-isolation})] 
or Drude [Eq.~(\ref{eqn:sd-drude})] isolation network 
at fixed temperatures $T = 0$ (main figure) 
and $T = 50$mK (inset), with $J_{int} (\omega) = 0$ corresponding to $R_0 \to \infty$.
In the limit of low temperatures $(k_B T/\hbar \omega_{01} \ll 1)$, the
relaxation time becomes 
\begin{equation}
T_1 (\omega_{01}) 
= 
\frac{M \omega_{01}}{J(\omega_{01})}. 
\end{equation}
From the main plot of 
Fig.~\ref{fig:four} several important points can be extracted. 
First, in the low frequency regime ($\omega_{01} \ll \Omega)$ 
the RL (Drude) and RLC environments 
produce essentially the same relaxation time $T_{1,RLC} (0) = T_{1,RL} (0) = T_{1,0} 
\approx (L_1/L)^2 R C_0$, because both systems are ohmic. 
Second, near resonance ($\omega_{01} \approx \Omega$), $T_{1,RLC}$ is 
substantially reduced 
because the qubit is resonantly coupled to its environment producing 
a distinct non-ohmic behavior.
Third, for ($\omega_{01} > \Omega$), $T_1$ grows very rapidly in the RLC case. 
Notice that for $\omega_{01} > \sqrt{2} \Omega$, the RLC relaxation time $T_{1,RLC}$ 
is always larger than $T_{1,RL}$.
Furthermore, in the limit of 
$\omega_{01} \gg {\rm max} \lbrace {\Omega, 2\Gamma} \rbrace$, 
$T_{1,RLC}$ grows with the fourth power of $\omega_{01}$ behaving as 
\begin{equation}
T_{1,RLC} 
\approx 
T_{1,0} 
\frac{\omega_{01}^4}{\Omega^4},
\end{equation}
while for $\omega_{01} \gg \Omega^2/2\Gamma$, $T_{1,RL}$ grows only with 
second power of $\omega_{01}$ behaving as 
\begin{equation}
T_{1,RL} 
\approx 
4 T_{1,0} 
\frac{\Gamma^2 \omega_{01}^2}{\Omega^4}
.
\end{equation}
Thus, $T_{1,RLC}$ is always much larger than
$T_{1,RL}$ for sufficiently large $\omega_{01}$. 
For parameters in the experimental range (Fig.~\ref{fig:four}), 
$T_{1,RLC}$ is two orders of magnitude 
larger than $T_{1,RL}$, indicating a clear advantage of the RLC environment 
shown in Fig~\ref{fig:one} over the standard ohmic RL environment.
Thermal effects are shown in the inset of Fig.~\ref{fig:four}, where $T = 50$mK 
is a characteristic experimental temperature~\cite{paik-08}. Typical values of $T_1$ at 
low frequencies vary from $10^{-5}$s at $T =0$ to $10^{-6}$s at $T = 50$mK, 
while the high frequency values remain essentially unchanged as 
thermal effects are not important for $\hbar \omega_{01} \gg k_B T$. 

\begin{figure}
\centerline{ \scalebox{0.65} {\includegraphics{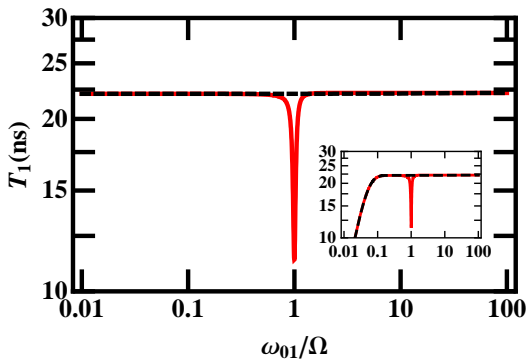}} } 
\caption{\label{fig:five}
(Color-online) $T_1$ (in nanoseconds) as a function of qubit frequency 
$\omega_{01}$.
The solid (red) curves describe a phase qubit with RLC isolation 
network (Fig.~\ref{fig:three}) with 
same parameters of Fig.~\ref{fig:two} except that $R_0 = 5000$~ohms. 
%%parameters $R = 50$~ohms, $L_1=3.9$nH, $L=2.25$pH, $C=2.22$pF, and 
%%qubit parameters $C_0=4.44$pF, $R_0 = 5000$~ohms and $L_0 = 0$. 
The dashed curves correspond to an RL isolation network with the same parameters of the RLC
network, except that $C = 0$. 
Main figure ($T =0$), inset ($T = 50$mK) with $\Omega = 141 \times 10^9$ rad/sec.}
\end{figure}

In the preceeding analysis we neglected the effect of the local environment by 
setting $Y_{int} (\omega) = 0$. As a result, the low-frequency value of $T_1$ is
substantially larger than obtained in experiment~\cite{martinis-02, paik-08}. By modeling
the local environment with $R_0 = 5000$~ohms and $L_0 = 0$, we obtain the $T_1$ 
versus $\omega_{01}$ plot shown in Fig.~\ref{fig:five}.
Notice that this value of $R_0$ brings $T_1$ to values close to $20$ns at $T = 0$.
The message to extract from Figs.~\ref{fig:four} and~\ref{fig:five} is 
that increasing $R_0$ as much as 
possible and increasing the qubit frequency $\omega_{01}$ from $0.1\Omega$ to 
$2\Omega$ at fixed
low temperature can produce a large increase in $T_1$. 

Having discussed the decoherence properties caused by environments 
with resonances for the case of superconducting qubits, 
we discuss next how the environment shifts (renormalizes) 
the qubit frequency for the same systems.

\section{Frequency renormalization}
\label{sec:frequency-normalization}

In this section, we discuss another effect of the 
environment on qubit properties, in addition
to dephasing and relaxation. The environment
also renormalizes (shifts) the qubit frequency 
$\omega_{01}$ by dressing the two-state system 
with environmental degrees of freedom. This is
similar to the physics of the Lamb shift or the Franck Condon effect.
In our case,  the transition frequency $E = \omega_{01}$ 
is renormalized according to $E_R = E + \delta E$, 
where $\delta E = -{\rm Im} R_{1010}$ in terms of the Redfield
rate tensor. The energy shift can be explicitly written as 
\begin{equation}
\label{eqn:lamb}
\delta E
=
\frac{1}{4\pi}{\cal P}
\int_0^{\infty} d\omega \frac{J(\omega)}{E^2-\omega^2}
\left[ E \coth(\beta\omega/2) - \omega\right],
\end{equation}
where ${\cal P}$ denotes the Cauchy principal value,
and $\beta = 1/k_B T$. 
Notice that $\delta E$ is analogous to the energy shift obtained
in second order perturbation theory, which collects all processes 
in which a virtual boson is emitted and reabsorbed, 
such that no trace is left in the environment. 
The integral in Eq.~(\ref{eqn:lamb}) 
can be calculated by extending the integration 
to the complete real axis, closing the countour 
in the upper complex plane, and applying the residue theorem. 

\subsection{Phase and flux qubits}
For phase and flux qubit discussed above, upon summation
over all residues of the relevant poles of the spectral density, we
arrive to~\cite{wilhelm-03}  
%
%%
%% resulting Matsubara  series. We end up with
%
%
%
%%
\begin{eqnarray}
\delta E
=
\frac{K}{2\pi}\frac{\Omega^3 E}{2i\Gamma} 
\sum_{\sigma = \pm }
\frac{\sigma}{E^2-(\sigma\Omega+i\Gamma)^2} \times \nonumber\\
\left[
G(\Gamma-i\sigma)
-{\rm Re} G(iE)-\pi\frac{\Gamma-i\sigma \Omega}{E}\right]
\label{eq:matsubara}
\end{eqnarray}
where $G(x)=\psi(1+\beta x/2\pi)+\psi(\beta x/2\pi)$ involves the
digamma function $\psi$, and $K = \alpha (\Phi_0/2\pi)^2$ for phase qubit 
[see Eq.~(\ref{eqn:spectral-density-isolation})], and 
$K = \alpha_1$ [see Eq.~(\ref{eqn:spectral-density})] for 
the flux qubit.     
%
%%
%
%
%%
%\textcolor{red} {*** I presume that the summation is over $\sigma = \pm$
%now, since you seem to have already summed up the poles of 
%the $\coth$ function. ***}
%%
%
Although, we use a compact notation involving
complex functions, the energy correction $\delta E$ is real.
Notice that $\delta E$ changes sign at
$\omega\simeq E$, leading to an upward shift (renormalization)
of $E$ if most of the spectral weight of $J(\omega)$ is above $E$ 
(corresponding to $E < \Omega$) whereas $E$ 
shifts downward in the opposite case. 
Physically, this corresponds to level repulsion between 
the spin and the oscillators in the environment. This result is
consistent with usual second-order perturbation theory for
energies. If the spectral weight $J(\omega)$ is concentrated at 
frequency $\omega = \Omega$, then the sign changes of $\delta E$ 
happens at $E \simeq \Omega$, leading to a rather
sharp structure in $\delta E (\Omega)$, and in $E_R (\Omega)$.

In the limit of low temperatures $(T \to 0)$, we can replace the function
$G$ appearing in Eq.~(\ref{eq:matsubara}) by an appropriate logarithm to find
\begin{equation}
\delta E
=
\frac{K}{2\pi}\frac{i\Omega^3 E}{\Gamma} 
\sum_{\sigma = \pm}
\frac{\sigma}{E^2-(\sigma \Omega + i\Gamma)^2}
\log\left(\frac{\Gamma-i\sigma\widetilde{\Omega}}{iE}\right),
\label{eq:br_renormal_full}
\end{equation}
where $\widetilde{\Omega} = \Omega-\Gamma^2/\Omega$. 
In the underdamped limit $\Omega \gg \Gamma$, one approximates the
logarithm by $\log|\Omega/E|-i\sigma\pi/2$ and rewrite the result as
$E_R = E + \delta E$,  with 
$\delta E = \delta E_\Omega + \delta E_{\rm res}$.  
The first term of $\delta E$ 
contains a logarithmic contribution which resembles the
scaling in the Ohmic case (with cutoff frequency $\widetilde\Omega$),
\begin{equation}
\delta E_\Omega
=
\frac{2}{\pi}J(E)\log
\left|\frac{E}{\widetilde{\Omega}}\right|.
\label{eq:br_renormal_ohm}
\end{equation}
where $\widetilde\Omega \approx \Omega$. 
%$\widetilde\Omega = \Omega -\Gamma^2/\Omega$. 
This contribution changes sign from an upward shift at 
$E > \widetilde\Omega$ to a
downward shift at $E < \widetilde\Omega $ as expected from the general
arguments described above. 

The other contribution to $\delta E$ takes into account the enormous
spectral weight of the resonance, and can be written as
\begin{equation}
\delta E_{\rm res}
=
J(E)
\frac{E^2-\widetilde{\Omega}^2}{\Omega\Gamma}.
\label{eq:br_renormal_peak}
\end{equation}
A term of this kind persists even in the absence of damping of the 
external oscillator. 
%
%%
%\textcolor{red}{*** what do you mean, when $\Gamma \to 0$? ***}
%%
%
This term vanishes linearly with $E$ at low energies, but
undergoes the expected sign change near resonance $E \approx \Omega$,
in which vicinity, a substantial renormalization also occurs.

\begin{figure} [htb]
\centerline{ \scalebox{0.85} {\includegraphics{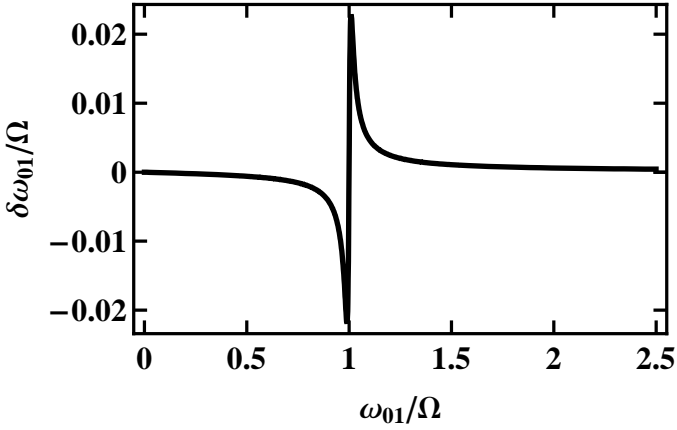}} } 
\caption{\label{fig:six}
Renormalization of energy splitting for the phase qubit with 
RLC isolation network 
(Fig.~\ref{fig:three}) for 
the parameters $R = 50$~ohms, $L_1=3.9$nH, $L=2.25$pH, $C=2.22$pF, 
and qubit parameters $C_0=4.44$pF, $R_0 = 5000$~ohms and $L_0 = 0$, $T=0$, and $\Omega = 141 \times 10^9$ rad/sec.}
\end{figure}

As an illustration of the qualitative results discussed
in this section, we show in Fig.~\ref{fig:six} 
the frequency shift (renormalization) 
of the phase qubit with RLC isolation network described
in Fig.~\ref{fig:three}.
We make the identification 
$E = \omega_{01}$ and $\delta E = \delta \omega_{01}$.
Near resonance $\omega_{01} \approx \Omega$, we find 
a frequency renormalization 
of about $2\%$ which is due to the term $\delta E_{res}$.

Next, we discuss briefly the frequency renormalization of charge qubits
due to environmental effects.

\subsection{Charge qubits}

For charge qubits with spectral density given by
Eq.~(\ref{eqn:spectral-density1}), the energy renormalization $\delta E$ 
obtained in Eq.~(\ref{eqn:lamb}) can be also calculated using complex 
integration techniques. In the low temperature limit $(T \to 0)$ there
are two contributions to 
\begin{equation}
\label{eqn:frequencyrenormalization2}
\delta E = \delta E_{\Omega} + \delta E_{res}.
\end{equation}
The first one is a resonant contribution
\begin{equation}
\delta E_{res} 
= 
e^2\frac{\Omega_2^2}{lc}
\frac{\Lambda_2
\left[
\pi + 2\arctan(2\Omega_2/\Gamma_2)
\right]}
{
\left(
\Lambda_2^2
+ \Gamma_2^2\Omega_2^2
\right)
}, 
\end{equation}
where the function 
\begin{equation}
\Lambda_2 
=
E^2-\Omega_2^2 +
\left(
\frac{
\Gamma_2}{2}
\right)^2 
\end{equation}
and the second one is a non-resonant contribution
\begin{equation}
\delta E_{\Omega} 
= 
e^2\frac{\Omega_2^2}{lc}
\frac{\Gamma_2\Omega_2
\log 
\left[
E^2/(\Gamma_2^2/4 + \Omega_2^2)
\right]}
{
\left(
\Lambda_2^2 + \Gamma^2\Omega_2^2
\right)
}
.
\end{equation}
Just like the phase and flux qubit, the 
frequeny renormalization term due to the spectral weight of the 
resonance $\delta E_{res}$ is much larger than the 
logarithmic contribution $\delta E_{\Omega}$. While $\delta E_{res}$
undergoes a sign change from negative to positive 
at $E =
\sqrt{\Omega_2^2 - (\Gamma_2/2)^2}$, $\delta E_{\Omega}$ undergoes a sign change at 
$\sqrt{\Omega_2^2 + (\Gamma_2/2)^2}$. 

Having concluded the discussion of the frequency renormalization (shift) of 
superconducting qubits due to coupling to environments with resonant
frequency $\Omega$, we discuss next the behavior of superconducting
qubits coupled to the same environments, when the 
qubit frequency $\omega_{01}$ is close to $\Omega$, where essentially
exact solutions are possible.

\section{Non-Markovian behavior near resonance}
\label{sec:non-markovian}

The Bloch-Redfield equations described in Eq.~(\ref{eqn:d-matrix})
capture the long time behavior of the density matrix, 
but can not describe the short time behavior
of the system in particular near a resonance 
condition $\omega_{01} \approx \Omega$, where
the environmental spectral density $J(\omega \approx \Omega)$ is very large. 
In this case, only the environmental modes with 
$\omega_{k} = \Omega$ couple strongly to the two-level system, 
like a two-level atom interacting with an electromagnetic field cavity mode 
that has a finite lifetime.
Thus, next we derive the time evolution of the state of the 
flux, phase and charge qubits when the qubit frequency is 
close to an environmental resonance.

First, we restrict the Hamiltonian described 
in Eqs.~(\ref{eqn:hamiltonian}) 
and~(\ref{eqn:system-bath-hamiltonian}) only to boson modes 
with $\omega_{k} \approx \Omega \approx \omega_{01}$.
When $\omega_{k} \approx \Omega \approx \omega_{01}$, the Hamiltonian 
shown in Eqs.~(\ref{eqn:hamiltonian}) 
and~(\ref{eqn:system-bath-hamiltonian}) can be solved
in the rotating wave approximation using the complete basis set of 
system-bath product states 
$\vert \psi_0 \rangle = \vert 0 \rangle_{\rm S} \otimes \vert 0\rangle_{\rm B}$;
$\vert \psi_1 \rangle = \vert 1 \rangle_{\rm S} \otimes \vert 0\rangle_{\rm B}$; 
$\vert \psi_k \rangle = \vert 0\rangle_{\rm S} \otimes \vert k \rangle_{\rm B}$, where
$\vert 0 \rangle_{\rm S}$ and $\vert 1 \rangle_{\rm S}$ are the states of the
qubit and $\vert k \rangle_{\rm B}$ are the states of the bath. Notice that the states 
$\vert 1 \rangle_{\rm S} \otimes \vert k\rangle_{\rm B}$ are absent in the 
basis set within the rotating wave approximation and that 
the state of the total system at any time is 
\begin{equation}
\label{eq:phi_exp}
\phi(t)=c_0\psi_0+c_1(t)\psi_1+\sum_{k\ne 0,1} c_k(t)\psi_k,
\end{equation}
with probability amplitudes $c_0$, $c_1(t)$, and $c_k(t)$. 
The amplitude $c_0$ is constant, while the amplitudes $c_1(t)$ and $c_k(t)$ are time dependent. 
Assuming that there are no excited bath modes at $t = 0$, we impose the
initial condition $c_k(0) = 0$, and use the normalization 
$\vert \phi (t) \vert^2 = 1$ to obtain the closed integro-differential equation
\begin{equation}
\dot{c}_1 (t) = - \int_0^t dt_1 f(t-t_1) c_1(t_1),
\end{equation}
where the kernel is the correlation function
$$
f(\tau) = \int_0^\infty d\omega J(\omega) \exp\left[\text{i}(\omega_{01}-\omega)\tau\right]
\nonumber
$$
directly related to the spectral density $J(\omega)$.
%
%%
%\textcolor{red}{*** What are the limits of integration here? 
%I think that it should be from $\omega = 0$ to $\omega \to %\infty$.
%***}
%%
%
The reduced density matrix 
\begin{equation}
\label{eq:rho_ex}
\rho(t)=
\left(
\begin{array}{cc}
|c_1(t)|^2  & c_1(t) c_0^*\\
c_1^*(t)c_0 & |c_0|^2+\sum_k|c_k(t)|^2
\end{array}
\right)
\end{equation}
is subject to the condition
${\rm Tr} \rho (t) = 1$, or more 
explicitly  
$|c_0|^2+\sum_k|c_k(t)|^2 = 1  -  |c_1(t)|^2 $, 
which shows that the time dynamics of $\rho(t)$ is fully determined
by $c_1 (t)$ for a specified value of $c_0$.

\subsection{Phase and Flux qubits}
For phase and flux qubita with spectral densities 
given by Eq.~(\ref{eqn:spectral-density-isolation})
and Eq.~(\ref{eqn:spectral-density}) respectively, 
we can rewrite the spectral density as
\begin{equation}
\label{eqn:sd-poles}
J_{\rm res} (\omega) =  
\frac{K\Omega^3}{4\text{i}\Gamma}
\sum_{\sigma=\pm 1}\frac{\sigma\omega}
{\omega^2-\left(\sigma\widetilde{\Omega} + \text{i}\Gamma\right)^2},
\end{equation}
where $K = \alpha (\Phi_0/2\pi)^2$ for phase qubit 
[see Eq.~(\ref{eqn:spectral-density-isolation}],
and 
$K = \alpha_1$ [see Eq.~(\ref{eqn:spectral-density})] for 
the flux qubit.
This reveals a resonance at 
$\omega = \widetilde \Omega$ with linewidth $\Gamma$, 
implying that any internal off-resonance contribution to 
$J(\omega)$, such as  
$J_{int} (\omega = \widetilde \Omega)$ 
in the case of the phase qubit coupled to 
an RLC environment, 
can be neglected for any non-zero 
value of the resistance $R_0$.
In this case, the spectral density $J(\omega)$ dominated
by the resonant contribution 
$J (\omega) \approx J_{iso} (\omega) \approx J_{\rm res} (\omega)$. 
It is very important to emphasize, that eventhough $J_{\rm res} (\omega)$
is rewritten explicitly in terms of its poles in the complex plane, 
a simple inspection of Eq.~(\ref{eqn:sd-poles}) 
shows that $J_{\rm res} (\omega)$ is real.

For this spectral density, we can now solve for $c_1(t)$ exactly and
obtain the closed form
$$
c_1(t)=
{\cal L}^{-1} 
\left\{
\frac{ (s+\Gamma - i\omega_{01} )^2 + \Omega^2 - \Gamma^2}
{s \left[ (s + \Gamma - i\omega_{01} )^2 + \Omega^2 - \Gamma^2 \right] - 
\kappa \Omega^4\pi i/\Gamma}
\right\}
$$
where ${\cal L}^{-1} \lbrace F (s) \rbrace$ is the inverse Laplace transform
of $F(s)$, and 
$\kappa = K \times (\Phi_0/2\pi)^2 \approx 1/(\omega_{01} T_{1,0})$.

\begin{figure}[htb]
\centerline{ \scalebox{0.7} {\includegraphics{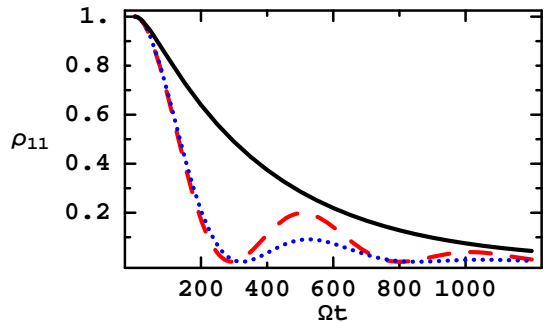}} } 
\caption{\label{fig:seven}
(Color-online) Population of the excited state of the dc-SQUID phase qubit 
(Fig.~\ref{fig:three}) as a function of time 
$\rho_{11}(t)$, with $\rho_{11} (t = 0) = 1$ for $R = 50$~ohms (solid black curve),
$350$~ohms (dotted blue curve), and $R = 550$~ohms (dashed red curve), and
$L_1=3.9$nH, $L=2.25$pH, $C=2.22$pF, $C_0=4.44$pF, $R_0 = \infty$ and $L_0 = 0$.}
\end{figure}

In Fig.~\ref{fig:seven}, we plot the density matrix element 
$\rho_{11} = \vert c_1 (t) \vert^2$ 
as a function of time for the dc-SQUID phase qubit 
described in Fig~\ref{fig:three}. The plot contains curves for
three different values of resistance, assuming that the qubit is 
initially (time $t =0$) in its excited state where $\rho_{11} (0) = 1$.
We consider the experimentally relevant weak dissipation limit 
of $\Gamma \ll \omega_{01} \approx \Omega$. Since $\Gamma = 1/(2CR)$ the width 
of the resonance in the spectral density shown in Eq.~(\ref{eqn:sd-poles}) is 
smaller for larger values of $R$. Thus, for large $R$, the RLC environment 
transfers energy resonantly back and forth to the qubit and induces
Rabi-oscillations with an effective time dependent decay rate 
$
\gamma(t) = - 2 {\cal R}\{ \dot c_1(t) / c_1(t) \}.
$

These environmentally-induced Rabi oscillations are a clear signature of 
the non-Markovian behavior produced by the RLC environment, and are completely 
absent in the RL environment because the energy from the qubits is quickly 
dissipated without being temporarily stored. In the RL environment the decay 
in time of $\rho_{11} (t)$ has the characteristic non-oscillatory Markovian behavior.
These environmentally-induced Rabi oscillations are generic features of 
circuits with resonances in the real part of the admittance. 
The frequency of the Rabi oscillations $\Omega_{Ra} = \sqrt{\pi \kappa \Omega^3/2\Gamma}$
is independent of the resistance 
since $\Omega_{Ra} \approx \Omega \sqrt{\pi L^2 C/L_1^2 C_0}$,
and has the value of 
$\Omega_{Ra} = 2\pi f_{Ra} \approx 360 \times 10^{6}$~rad/sec in 
Fig.~\ref{fig:seven}. This effect is similar to the so-called circuit 
quantum electrodynamics which has been of great experimental interest 
recently~\cite{wallraff-04, chiorescu-04, sillanpaa-07} 
%{\textcolor{red}{in the context of charge qubits}}.
%
%%

Next, we discuss briefly the non-Markovian dynamics for charged qubits.

\subsection{Charge qubits}

For charge qubits, the spectral density in 
Eq.~(\ref{eqn:spectral-density1}) can be rewritten 
in terms of its poles in the complex plane as 
\begin{equation}
J(\omega) = \frac{e^2\Omega_2}{ilc}\sum_{\sigma=\pm 1}
\frac{\sigma}{\omega-	\Omega_2+\sigma i\Gamma_2/2}.
\end{equation}
Again, we can solve for $c_1(t)$ exactly, which is now given by
\begin{equation}
\label{eqn:c1-charge-qubit}
c_1(t) = c_1(0) e^{-\Gamma_2 t/4} \left( \cosh 
\frac{\Delta t}{2} + \frac{\Gamma_2}{2\Delta}\sinh \frac{\Delta t}{2}\right)
\end{equation}
where $\Delta = \sqrt{\Gamma_2^2/4 -8\pi e^2 \Omega_2/lc}$.   
Notice that the frequency $\Delta$ can be complex, so it is convenient
to rewrite it as 
\begin{equation}
\Delta 
= 
\frac{\Gamma_2}{2}
\sqrt{
1 
- 
\left(
\frac{2\Omega_c}{\Gamma_2}
\right)
\left(
\frac{2\Omega_2}{\Gamma_2}
\right),
}
\end{equation}
where $\Omega_c = 8\pi e^2/\ell c$.
In the underdamped case where $\Omega_2 \gg \Gamma_2/2$, 
provided that $2 \Omega_c/\Gamma_2$ is not too small,
$\Delta$ becomes purely imaginary:
\begin{equation}
\Delta 
\approx
i \sqrt{\Omega_c \Omega_2}.
\end{equation}
Since $\cosh(i\theta) = \cos(\theta)$ and
$\sinh(i\theta) = i \sin(\theta)$, where
$\theta = (\sqrt{\Omega_c \Omega_2})t/2$ in
the present case, it becomes clear that an
oscillatory decay of the state emerges
in the underdamped regime. In this regime, 
the period of oscillation is just 
$\tau = 4\pi/\sqrt{\Omega_c \Omega_2}$.
%
%%Clearly, in the underdamped case, oscillatory decay of the state emerges.
%
%
%%

Having completed our discussion of environmentally-induced Rabi oscillations and its
markedly non-markovian characteristic, we are ready to conclude.

\section{Conclusions}
\label{sec:conclusions}

In conclusion, 
we analyzed decoherence effects in qubits coupled to environments 
containing resonances in their spectral function,
and we identified a crucial role played by the design 
of isolation circuits on decoherence properties. 
Furthermore, we found that the decoherence time of qubits can 
be two orders of magnitude larger than their typical low-frequency ohmic-regime, 
provided that the frequency of the qubit is about two times larger 
than the resonance frequency of the environmental resonance (isolation
circuit in the phase qubit case).
We also studied the frequency renormalization (shift) of the
qubit described by a two-level system due to dressing of the  
energy levels by the environmental degrees of freedom.
We found that the frequency shift changes sign across the resonance 
frequency and is largest at resonance (about $2 \%$).
Lastly, we showed that when the qubit frequency is close to a resonance
of the environment, the non-oscillatory
Markovian decay of the excited state population of the qubit, gives in to an oscillatory
non-Markovian decay, as the qubit and its environment self-generate Rabi 
oscillations of characteristic time scales shorter than the decoherence time.
In particular, we discussed as a concrete example the decoherence properties of
a dc-SQUID phase qubit coupled with an environmental RLC circuit possesing a 
resonance in its spectral function, where numbers compatible with current experiments
were used to estimate the environmental effects on decoherence properties~\cite{mitra-08,paik-08}.
%
%%
%\textcolor{red}{*** Perhaps one should cite the MD experiments here. ***}
%%
%
%
%%
\begin{acknowledgements}
K. Mitra and C. J. Lobb would like to 
acknowledge support from the National Science Foundation (DMR-0304380) 
and NSA, through the Laboratory of Physical Sciences, and
C. A. R. S{\'a} de Melo would like to acknowledge support
from the National Science Foundation (DMR-0709584) and 
the Army Research Office (W911NF-09-1-0220).
\end{acknowledgements}
%%
%

%
%%
%\textcolor{red}{*** Since there is no space problem, all the references should
%appear with the names of all the authors. *** This is the standard recommendation
%of PRB, if you do not do it now, they will probably ask you do it before the paper
%gets published. ***}

\end{document}